\documentclass[aps,prb,twocolumn,superscriptaddress,showpacs]{revtex4}

\usepackage{graphics}
\usepackage{epsfig}
\usepackage{dcolumn}
\usepackage{bm}
\usepackage{amsmath}
\usepackage{amsfonts}
\usepackage{latexsym}
\usepackage{amssymb}
\usepackage{color}

\usepackage{hyperref}

\newcommand{\g}[1]{{\bf #1}}

\begin{document}

\title{Seebeck effect in the graphene-superconductor junction}

\author{Marcin M. Wysoki\'nski}
\email{marcin.wysokinski@uj.edu.pl}
\affiliation{Marian Smoluchowski Institute of Physics$,$ Jagiellonian University$,$ Reymonta 4$,$ PL-30-059 Krak\'ow$,$ Poland}
\author{Jozef Spa\l ek}
\email{ufspalek@if.uj.edu.pl}

\affiliation{Marian Smoluchowski Institute of Physics$,$ Jagiellonian University$,$ Reymonta 4$,$ PL-30-059 Krak\'ow$,$ Poland}
\affiliation{Faculty of Physics and Applied Computer Science$,$ AGH University of Science and Technology$,$ Reymonta 19, PL-30-059 Krak\'ow, Poland}

\date{\today}

\begin{abstract}
Thermopower of graphene-superconductor 
(GS) junction is analyzed within the
extended Blonder-Tinkham-Klapwijk formalism. 
Within this approach we have also calculated
the temperature dependence of 
the zero-bias conductance for GS junction. 
Both quantities reflect quasi-relativistic nature of 
massless Dirac fermions in graphene. Both, the linear
and the non-linear regimes are considered.
\end{abstract}

\pacs{74.45.+c, 73.23.Ad, 65.80.Ck}

\maketitle
\section{Introduction}
Graphene is one of the most remarkable new materials. 
Not only has its discovery \cite{novoselov2005} 
violated in same sense Landau's theory of the thermodynamical instability of 
a two-dimensional structure\cite{landau1980}, but also due to the peculiar band structure 
it has provided us with an invaluable opportunity to test relativistic quantum electrodynamics 
in the desktop laboratory \cite{beenakker2008}. For that reason, much effort has been 
put into understanding of the phenomena associated with this material.

Since the graphene-based devices are usually considered as a mesoscopic systems, 
the Landauer approach is widely utilized to study the ballistic transport in them
\cite{tworzydlo2006,snyman2007,xing2009}. Even though this approach
 does not account in the simplest form for all the features 
of the material, it provides a good overall 
description of the electric transport.
This approach was extended by Blonder, Tinkham, and Klapwijk\cite{blonder1982} (BTK) 
to the case of standard normal metal-superconductor junction (NS), 
and in this manner yielded a very good description of 
experimental data\cite{blonder1983}. Their method has been
widely used for different specific situations
\cite{kaczmarczyk2011,PhysicaC2011,wysokinski2012,annunziata2011,bardas1995,
mortensen1999, tanaka1995,lukic2007,hirsch1994} and finally adapted for 
graphene-superconductor hybrid systems 
\cite{beenakker2006,linder2008,takehito2008,zhang2008, titov2006}.
One of the most peculiar properties predicted in such systems is the specular 
Andreev reflection and deviations of the conductance spectra\cite{beenakker2006} from
those predicted by BTK for normal metals\cite{blonder1982}.

Landauer formalism has also been successfully adapted for thermoelectrical
transport in mesoscopic devices\cite{sivan1986,butcher1990,houten1992}.
In the case of standard NS junctions BTK formula, also 
turned out to be useful technique for 
predicting effects concerning thermal properties 
of electric and heat currents 
\cite{bardas1995,wysokinski2012,hirsch1994,devyatov2010}.
This method has also been used for the graphene-based superconducting 
hybrid structures for obtaining the thermal conductance
\cite{titov2007,takehito2008,salehi2010,2salehi2010}.
However, the thermopower has not been studied so far. 
This topic is addressed in this article.

In this work we provide systematic study of the effect of the temperature 
on the charge current in the superconducting graphene junction (GS) using
a generalized BTK formalism for the specific case of graphene. 
We present results concerning the temperature dependence of the 
zero-bias conductance and the Seebeck coefficient in the linear regime. 
For the sake of completeness, we also discuss the non-linear thermopower.

The paper is organized as follows. In Sec. \ref{model} (and in Appendix \ref{appA}), 
we present briefly a generalized BTK approach 
for the charge current through the GS junction. 
The linear transport coefficients are
discussed, in particular the 
zero-bias conductance and the thermopower. 
We also briefly comment on the effect 
of non-linear corrections on the Seebeck coefficient. 
Finally, we conclude in the Sec. \ref{conclusions}. 

\section{Model}
\label{model}
\begin{figure}[b]
\begin{center}
\includegraphics[width=0.36\textwidth]{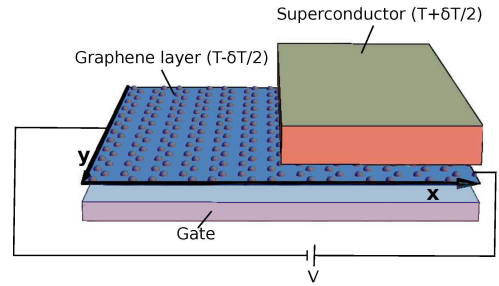}
\caption{(Color online) Proposed schematic, 
experimental setup considered in our modeling.}
\label{setup}
\end{center}
\end{figure}

We consider a ballistic limit for graphene based junction composed of the normal region
 and induced by means of the proximity effect superconducting region (cf. Fig.\ref{setup}). 
For the description of the unconventional quasiparticle states we utilize
Dirac - Bogoliubov - de Gennes equations for the two-dimensional (2D)
 sheet of graphene in the form\cite{degennes1966,beenakker2006}
{\small
 \begin{equation}
 \begin{pmatrix}
  H_j - E_F \g{1}&\Delta  \\
  \Delta^\dagger & E_F \g{1}-H_j
 \end{pmatrix}
\begin{pmatrix}
  u  \\
  v 
 \end{pmatrix}
=\epsilon \begin{pmatrix}
u  \\
  v
 \end{pmatrix},
\label{bdg}
\end{equation}}
where the index $j$ can be either + or - what refers to the 
two inequivalent valleys K and K' in the Brillouin zone.
The single particle Hamiltonian is given by
\begin{equation}
 H_{\pm}=-i\hslash v_F(\sigma_x\partial_x\pm \sigma_y\partial_y)+U,
\end{equation}
where $v_F$ is the energy independent Fermi velocity for the graphene, and 
\{$\sigma_i$\} denote respective Pauli matrices.
Because of the valley degeneracy one can effectively do the calculation for the
 one valley only. We assume that in the geometry, where the interface is determined
 by the $y$-axis, the pair potential with the s-wave symmetry changes 
step-like in the x-axis direction,
\begin{equation}
 \Delta(\g{r},T)=\left\{ 
  \begin{array}{l l}
0 \ \ \ \ \ \ \ \ \ \ \ \  \ \ \ \ x<0,\\
\Delta(T)e^{i\phi} \ \ \ \ \ \ x>0,
\end{array}\right.
\end{equation}
where the temperature dependence of the gap 
function can be deduced from 
the usual BCS theory\cite{ketterson1999} and is taken in the following form 	
\begin{equation}
\frac{\Delta(T)}{\Delta_0}=\tanh\left(\sqrt{1.76\cdot\sqrt{\frac{T_c}{T}-1}}\right).
\end{equation}

The BCS theory of the superconductivity based on the requirement 
that the coherence length is large when compared to the Fermi wavelength. 
Under that condition we can assume that the additional potential $U$ has the
form,
\begin{equation}
 U(\g{r})=\left\{ 
  \begin{array}{l l}
0 \ \ \ \ \ \ \ \ \ \ \ \ x<0,\\
-U_0 \ \ \ \ \ \ \ \ x>0,
\end{array}\right.
\end{equation}
with large $U_0$ ($E_F+U_0>>\Delta_0$) and for the simplicity we define $E_F'=E_F+U_0$.
In numerical calculations we set $E_F'=1000\Delta_0$.

In the spirit of the BTK scheme (by matching the wave functions at the boundary $(x=0)$),
we obtain expressions for the amplitudes
of the Andreev hole reflection (AR) ($a(\epsilon,\theta)$) and the normal reflection 
($b(\epsilon,\theta)$)
of the incident electron - see Appendix \ref{appA} for details of the method.
Note that there is no intrinsic barrier at the GS junction, and thus the
Fermi vector mismatch is the source of the normal reflection. 
The transmission probability, averaged over the angles,
takes the following form \cite{blonder1982,mortensen1999},
{\small
\begin{equation}
\mathcal{T}(\epsilon)=\int_{-\pi/2}^{\pi/2} d\theta
\frac{\cos\theta}{2}\left(1-|b(\epsilon,\theta)|^2+
\frac{\mbox{Re} [e^{i\theta_A}]}{\cos\theta}|a(\epsilon,\theta)|^2\right).
\end{equation}}
 The BTK formalism combined with the specific transmission
probability derived for graphene, defines charge current 
through the GS interface as\cite{blonder1982,beenakker2006},
\begin{equation}
\begin{split}
I_{e} =\frac{4e}{h}\int_{-\infty}^{\infty}d\epsilon\ N(\epsilon)
\mathcal{T}(\epsilon)
\left(f^G(\epsilon-eV)-f^S(\epsilon)\right),
\label{BTK}
\end{split}
\end{equation}
where $f^G$, $f^S$ are the Fermi distribution functions for the normal (G)
 and the superconducting (S) region of GS junction respectively, and 
\begin{equation}
N(\epsilon)=\frac{|E_F+\epsilon|W}{\pi\hslash v_F},
\end{equation}
is the energy dependent number of transverse modes
in the graphene sheet of width $W$\cite{beenakker2006}. However, 
formula (\ref{BTK})
is not always accurate. The additional assumption is needed, 
that for each mode carrying incident electron, having energy 
$E_F+\epsilon$ and being Andreev reflected there is always
enough modes at the level $E_F-\epsilon$ for this 
process to happen. However, away from perfect Andreev 
reflection regime ($|a(\epsilon)|^2\neq1$), as in our case,
formula (\ref{BTK}) remains rigorous.

The quantity describing thermoelectric properties
of the system is thermopower, or Seeback coefficient ($S$) measuring
the voltage driving to zero the current flowing in response to the temperature difference, namely
\begin{equation}
 S\equiv-\left(\frac{V}{\delta T}\right)_{I_e=0}.
\label{thermo}
\end{equation}

\section{Charge Transport}
\label{transport}
\subsection{Linear regime}
Expansion of the Fermi functions in the normal and superconducting regions
 to the first (linear) order in both the bias and the temperature difference, with the average 
temperature $T$, i.e.,
\begin{equation}
 \begin{gathered}
 f^G\equiv f_{T-\delta T/2}(\epsilon-eV)\simeq f_T(\epsilon)-eV\frac{\partial f}{\partial \epsilon}+\frac{\delta T}{2T}\epsilon\frac{\partial f}{\partial \epsilon}, \\
f^S\equiv f_{T+\delta T/2}(\epsilon)\simeq f_T(\epsilon)-\frac{\delta T}{2T}\epsilon\frac{\partial f}{\partial \epsilon},
 \end{gathered}
\end{equation}
enables to decouple Eq.(\ref{BTK}) in the form,
\begin{equation}
\begin{gathered}
 I_e=GV+I_e^T\delta T.
\label{onsager}
\end{gathered}
\end{equation}
The resulting from it the linear transport coefficients can be thus rewritten in the 
following closed forms,
\begin{equation}
 \begin{gathered}
  G=-\frac{4e^2}{h}\int_{-\infty}^{\infty}d\epsilon\ 
\frac{\partial f}{\partial\epsilon}N(\epsilon)\mathcal{T}(\epsilon),\\
 I_e^T=\frac{4e}{hT}\int_{-\infty}^{\infty}d\epsilon\
\frac{\partial f}{\partial\epsilon}\epsilon N(\epsilon)\mathcal{T}(\epsilon).
\label{linearcoef}
 \end{gathered}
\end{equation}
The temperature gradient and the bias are set as positive with respect to $x$ coordinate.
For the temperature in the system approaching zero, the expression for the 
electric conductance (G) reduces to the well-known BTK zero-bias 
conductance formula\cite{blonder1982}
\begin{equation}
G_{T\rightarrow0}=\frac{4e^2}{h}N(0)\mathcal{T}(0).
\label{G}
\end{equation} 

\begin{figure}[t]
\begin{center}
\includegraphics[width=0.45\textwidth]{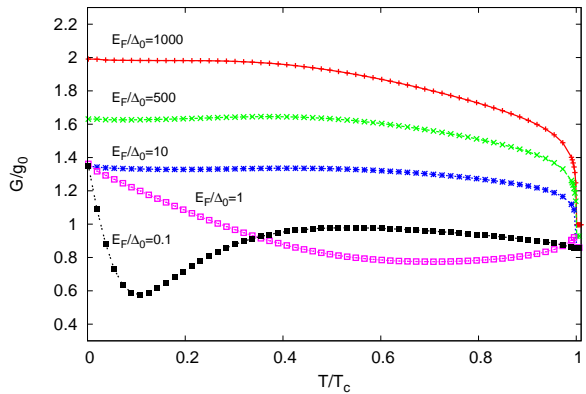}
\caption{(Color online) Normalized zero-bias conductance as a function of temperature for 
various Fermi vector mismatch.}
\label{linearG}
\end{center}
\end{figure}

In Fig.\ref{linearG} we have plotted the zero-bias differential 
conductance (calculated from Eq.(\ref{linearcoef})) as a function 
of temperature, normalized by the
ballistic conductance $g_0$ with having $N$ transverse modes in a 
sheet of graphene of width W and given by
\begin{equation}
 g_0=\frac{4e^2}{h}\int_{-\infty}^{\infty}d\epsilon\ 
\frac{\partial f}{\partial\epsilon}N(\epsilon).
\end{equation}
For relatively high position of the Fermi level in graphene 
(roughly $E_F\gtrsim5\Delta_0$), the influence of the increasing 
temperature (up to the approximately 
$T/T_c=0.5$) on the conductance is almost 
negligible (cf. Fig. \ref{linearG}). The origin of this behavior
is strictly connected with the relatively slow variation of the 
transmission probability $\mathcal{T}(\epsilon)$ for the
subgap energies in this temperature regime, and thus does not
differ qualitatively from the standard NS case\cite{blonder1982}.
In the low doping regime ($E_F\lesssim5\Delta_0$), 
the linear differential conductance as a function of the temperature 
vastly differs from  the standard 
NS case and reflects the specific electronic nature of 
graphene, as well as the impact of a crossover 
from the retro to the specular AR limit.

The non-trivial behavior of the Dirac fermions in graphene has also its influence on
the linear thermopower. From Eq. (\ref{thermo}) and (\ref{onsager})
formula for this quantity reads 
\begin{equation}
 S=\frac{I_e^T}{G}=-\frac{1}{k_BT}\frac{\int_{-\infty}^{\infty}d\epsilon\ \epsilon|\epsilon+E_F|
\frac{\partial f}{\partial\epsilon}\mathcal{T}(\epsilon)}{ 
\int_{-\infty}^{\infty}d\epsilon\ |\epsilon+E_F|
\frac{\partial f}{\partial\epsilon}\mathcal{T}(\epsilon)}\ \frac{k_B}{e}.
\end{equation}

\begin{figure}[t]
\begin{center}
\includegraphics[width=0.44\textwidth]{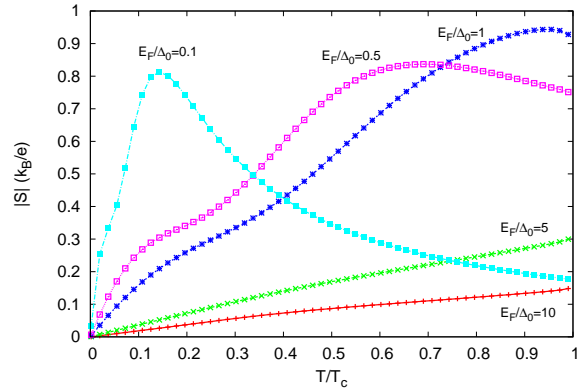}
\caption{(Color online) Linear 
thermopower as a function of temperature in the system for 
various Fermi vector mismatch.}
\label{linearTEP}
\end{center}
\end{figure}

Results obtained by the numerical integration are presented in the Fig. \ref{linearTEP}.
The low temperature regime differs from the one obtained for the NS junction
\cite{wysokinski2012}. Contrary to the NS case, for the graphene-based 
structure the thermopower does not vanish for non-zero temperatures.
This is due to the relativistic nature of charge carriers in graphene, 
where AR does not vanish even for the high effective barrier 
in our case effective barrier is only due to the Fermi level mismatch) 
for the subgap energies  
(for the normal incidence AR happens always with certainty\cite{beenakker2006}).
Furthermore, in the low doping regime
($E_F\lesssim\Delta_0$) the Seebeck coefficient for the GS junction 
is around one order of magnitude larger than even for high 
effective barrier value (which as a phenomenological parameter, can incorporate
also the Fermi velocity mismatch\cite{blonder1983}) in the NS case\cite{wysokinski2012}.

In this regime, we observe a clear maximum in the temperature
dependence of the Seebeck coefficient in the superconducting state. 
The maximum roughly corresponds to the minimum in the 
zero-bias conductance as a function of temperature (cf. Fig.\ref{linearG}).
The source of the significant enhancement 
of the thermopower in the low doping regime ($E_F\lesssim\Delta_0$)
should be understood as an effect of being partly in 
the specular AR limit, where the transmission probability spectrum
drops to zero for energy corresponding to the Fermi level position\cite{beenakker2006}. 
This feature affects the zero-bias conductance as a function of temperature and in turns
is responsible for the thermopower increase in the specular AR limit.
Therefore in this regime, the thermopower is 
large and reaches values up to $1 k_B/e$.
This suggests that there is a potential for
application of this setup for cooling of various nanostructures.

\subsection{Effect of the non-linearity}
\begin{figure}[t]
\begin{center}
\includegraphics[width=0.45\textwidth]{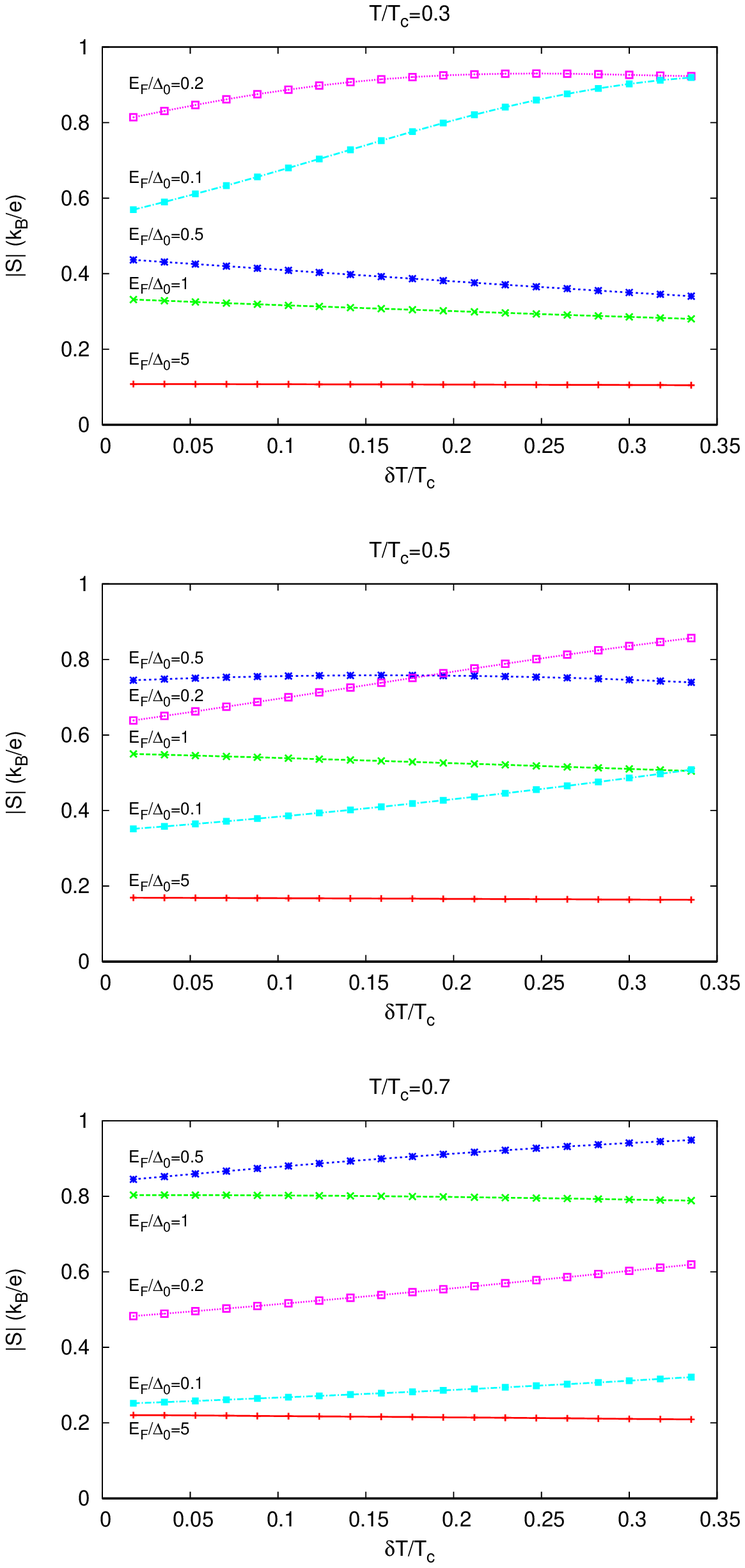}
\caption{(Color online) Nonlinear thermopower as a 
function of the temperature gradient set over the junction
for various Fermi energies in graphene. The net
temperature in the system is marked above each plot.}
\label{nonlinear}
\end{center}
\end{figure}

We have also studied numerically the effect of the non-linearity in our system.
The thermopower in the non-linear regime can be calculated as a ratio between
 the bias voltage and the temperature gradient, when no charge 
current is flowing (c.f. Eq.(\ref{thermo})). The results in the non-linear 
regime are presented in Fig. \ref{nonlinear}.
We have found that the non-linearity influences 
the thermopower in a not systematical manner with changing
Fermi level position and the average temperature of the system. 
However the change is not as dramatic as in 
the NS case\cite{wysokinski2012} and is almost unnoticeable in the
doped regime $E_F\gtrsim5\Delta_0$.

 \section{Conclusions}
\label{conclusions}
In this work we have analyzed the thermoelectric charge transport in the graphene 
junction consisting of the normal and the superconducting parts. In the linear 
regime we have calculated the temperature dependence of the zero-bias conductance and 
the thermopower. We have found deviations of these quantities 
from the standard normal metal-superconductor junction case 
that are caused by the relativistic nature of electrons in graphene. 
In the specular Andreev reflection regime 
Seebeck coefficient is strongly enhanced for specific temperatures.

We have also studied the effect of non-linearity on the thermopower
 and we have found that for a high Fermi level positions  
($E_F\gtrsim\Delta_0$) it stays almost unaffected and in the low 
Fermi level regime is noticeably enhanced with the increase of the 
temperature gradient.

\section*{Acknowledgements:} 
The authors greatly appreciate 
the stimulated discussion with Adam Rycerz, Jan Kaczmarczyk and Marcin Abram.
 The work has been partially supported by the Foundation for Polish Science (FNP) 
under the TEAM program. We also acknowledge the Grant MAESTRO from the National 
Science Center (NCN).

\appendix
\section{BTK for graphene}
\label{appA}
The wave function in the normal part of graphene (NG) $\psi_N$ and in the superconducting
region (SG) $\psi_S$, look respectively as follows
\begin{equation}
\begin{gathered}
 \psi_N=\psi_N^{e+}+ b\psi_N^{e-}+ a\psi_N^{h-}\\
\psi_S=c\psi_S^{e+}+d\psi_S^{h+},
\end{gathered}
\end{equation}
where the superscripts $e$, $h$ refers to electron and hole in 
NG and electronlike and holelike
excitation in SG, and the superscripts + and - to right 
and left moving particle respectively.

Spinors resulting from Eq.(\ref{bdg}) are expressed in the similar manner as 
in the Ref. \onlinecite{linder2008}, i.e., in the form
\begin{equation}
\begin{gathered}
 \psi_N^{e\pm}=[1,\pm e^{\pm i\theta},0,0]^Te^{\pm ik^ex\cos\theta},\\
 \psi_N^{h-}=[0,0,1,e^{-i\theta_A}]^Te^{-ik^hx\cos\theta_A},\\
\psi_S^{e+}=[u,u e^{i\theta^e_S},ve^{-i\phi},ve^{i(\theta^e_S-\phi)}]^Te^{iq^ex\cos\theta^e_S},\\
\psi_S^{h-}=[v,-v e^{-i\theta^h_S},ue^{-i\phi},-ue^{-i(\theta^h_S+\phi)}]^Te^{-iq^hx\cos\theta^h_S},
\end{gathered}
\end{equation}
where for the sake of clarity we do not include phase factor $e^{ik_yy}$ 
since it corresponds to conservation of momentum in $\bf{\hat y}$ direction. 
The corresponding wave vectors are defined as follows,
\begin{equation}
 k^{e(h)}=\frac{\epsilon+(-)E_F}{\hslash v_F}, \ \ \  
q^{e(h)}=\frac{E'_F+(-)\sqrt{\epsilon^2-\Delta^2}}{\hslash v_F},
\end{equation}
and the coherence factors are given by
\begin{equation}
 u=\sqrt{\frac{1}{2}\left(1+\frac{\sqrt{\epsilon^2-\Delta^2}}{\epsilon}\right)}, \ \
v=\sqrt{\frac{1}{2}\left(1-\frac{\sqrt{\epsilon^2-\Delta^2}}{\epsilon}\right)}.
\end{equation}

The conservation of momentum at the interface and along
 $\bf{\hat y}$ direction enables us to obtain mutual 
relations for the specific angles, namely
\begin{equation}
 k^e\sin\theta=k^h\sin\theta_A=q^e\sin\theta^e_S=q^h\sin\theta^h_S.
\end{equation}

The system must also satisfy the continuity condition at the interface, 
  $\psi_{\sigma L}(0)=\psi_{\sigma R}(0)$. The Hamiltonian is linear
therefore is no need in matching derivatives.
The resulting wave function amplitudes take the form,
\begin{widetext}
\begin{equation}
 \begin{gathered}
a(\epsilon,\theta)=\frac{2\cos\theta(e^{-i\theta_S^h}+e^{i\theta_S^e})uv}{
(e^{-i\theta_A}+e^{-i\theta_S^h})(e^{-i\theta}+e^{i\theta_S^e})u^2-
(e^{-i\theta}-e^{-i\theta_S^h})(e^{-i\theta_A}-e^{i\theta_S^e})v^2},\\
b(\epsilon,\theta)=\frac{2\cos\theta[(e^{i\theta_S^e}-e^{-i\theta_A})v^2+(e^{-i\theta_S^h}+e^{-i\theta_A})u^2]
}{(e^{-i\theta_A}+e^{-i\theta_S^h})(e^{-i\theta}+e^{i\theta_S^e})u^2-
(e^{-i\theta}-e^{-i\theta_S^h})(e^{-i\theta_A}-e^{i\theta_S^e})v^2}-1.
 \end{gathered}
\end{equation}
\end{widetext}

\end{document}